\documentclass[10pt]{IEEEtran}
\IEEEoverridecommandlockouts
\usepackage{dsfont}
\usepackage{algorithm}
\usepackage{algpseudocode}
\usepackage{listings}
\usepackage{amsmath}
\usepackage{mathtools}
\usepackage{listings}
\usepackage{lstlinebgrd}
\usepackage{tipa}
\usepackage[super]{nth}
\usepackage{subfig}
\usepackage[noadjust]{cite}
\usepackage[font={small,bf}]{caption}
\usepackage{listings, multicol, caption}
\usepackage{lstlinebgrd}
\usepackage{lipsum}
\usepackage{multirow}
\usepackage{paralist}
\usepackage{dsfont}
\usepackage{enumitem}

\def\BibTeX{{\rm B\kern-.05em{\sc i\kern-.025em b}\kern-.08em
    T\kern-.1667em\lower.7ex\hbox{E}\kern-.125emX}}
\begin{document}

%\title{Applying Convolutional Neural Networks (CNNs) to Branch Prediction by Modeling Global History}

%% \title{Improving Branch Prediction by Modeling 
%%   Global Branch Histories with Convolutional Neural Networks}

\title{Improving Branch Prediction By Modeling Global History with Convolutional Neural Networks}

\author{\IEEEauthorblockN{Stephen J. Tarsa, Chit-Kwan Lin, Gokce Keskin, Gautham Chinya, Hong Wang}\\
\IEEEauthorblockA{\textit{Intel Corporation} \\
Santa Clara, CA, USA \\
\{stephen.j.tarsa, chit-kwan.lin\}@intel.com}\vspace{-0.5cm}}
%% \and
%% \IEEEauthorblockN{2\textsuperscript{nd} Given Name Surname}
%% \IEEEauthorblockA{\textit{dept. name of organization (of Aff.)} \\
%% \textit{name of organization (of Aff.)}\\
%% City, Country \\
%% email address}
%% \and
%% \IEEEauthorblockN{3\textsuperscript{rd} Given Name Surname}
%% \IEEEauthorblockA{\textit{dept. name of organization (of Aff.)} \\
%% \textit{name of organization (of Aff.)}\\
%% City, Country \\
%% email address}
%% \and
%% \IEEEauthorblockN{4\textsuperscript{th} Given Name Surname}
%% \IEEEauthorblockA{\textit{dept. name of organization (of Aff.)} \\
%% \textit{name of organization (of Aff.)}\\
%% City, Country \\
%% email address}
%% \and
%% \IEEEauthorblockN{5\textsuperscript{th} Given Name Surname}
%% \IEEEauthorblockA{\textit{dept. name of organization (of Aff.)} \\
%% \textit{name of organization (of Aff.)}\\
%% City, Country \\
%% email address}
%% \and
%% \IEEEauthorblockN{6\textsuperscript{th} Given Name Surname}
%% \IEEEauthorblockA{\textit{dept. name of organization (of Aff.)} \\
%% \textit{name of organization (of Aff.)}\\
%% City, Country \\
%% email address}
%% }
\maketitle

\begin{abstract}
CPU branch prediction has hit a wall---existing techniques achieve near-perfect accuracy 
on 99\% of static branches, and yet the mispredictions that remain hide major performance 
gains.
%Even as existing techniques
%achieve near-perfect accuracy on 99\% of static branches, the
%remaining mispredictions hide major performance gains. 
In a companion
report, we show that a primary source of mispredictions is a handful of
systematically hard-to-predict branches (H2Ps), 
e.g. just 10 static instructions per SimPoint phase in SPECint 2017. 
The lost opportunity posed by these mispredictions is significant 
to the CPU: 14.0\% in instructions-per-cycle (IPC) on Intel SkyLake and 37.4\% IPC 
when the pipeline is scaled
four-fold, on par with gains from process technology.  However, up to 80\%
of this upside is unreachable by the best known branch predictors,
even when afforded exponentially more resources.
% @@@ckl I revised the percentages above (originally: 18.5\% and
% 55.3\%) to reflect just H2Ps because the subject of this paragraph
% is just H2Ps, not all of the mispredictions.

New approaches are needed, and machine learning (ML) provides a
palette of powerful predictors. A growing body of work has shown that
ML models are deployable within the microarchitecture to optimize
hardware at runtime, and are one way to customize CPUs post-silicon by
training to customer applications.  We develop this scenario for
branch prediction using convolutional neural networks (CNNs) to boost
accuracy for H2Ps.  Step-by-step, we (1) map CNNs to the global
history data used by existing branch predictors; (2) show how CNNs
improve H2P prediction in SPEC 2017; (3) adapt 2-bit CNN inference to
the constraints of current branch prediction units; and (4) establish
that \emph{CNN helper predictors} are reusable across application executions
on different inputs, enabling us to amortize offline
training and deploy ML pattern matching to improve IPC.
\end{abstract}

%% \begin{IEEEkeywords}
%% component, formatting, style, styling, insert
%% \end{IEEEkeywords}

\section{Introduction}
\label{sec:intro}
{\let\thefootnote\relax\footnote{\hrule \vspace{0.1cm} \emph{2nd ISCA International Workshop on AI Assisted Design for Architecture (AIDArc), June 2019, Phoenix, AZ, USA}}}
CPU branch predictors enable speculative execution and are a critical
tool for hiding latency in out-of-order cores. They work by inferring
the unresolved direction of a branch instruction when it is fetched,
based on a model trained to previously observed directions. Today,
branch prediction units (BPUs) perform both prediction and training
\emph{online} within a CPU's front-end, as an application runs. Though
tightly constrained, e.g. in storage and latency, existing predictors
achieve $>$99\% accuracy on 99\% of static branch instructions in SPEC
2017~\cite{fog2018microarchitecture, seznec_cbp16, lintarsa2019}.

However, the mispredictions that remain hide a major performance upside.  Our
data~\cite{lintarsa2019} shows that a small number of static branch 
instructions, just 10 on average per SPECint 2017 SimPoint phase, are systematically
mispredicted. Improving accuracy on these hard-to-predict branches (H2Ps) 
would boost instructions per cycle (IPC) on an Intel SkyLake core up-to 14.0\%, and up-to 37.4\% on a projected future CPU pipeline with width
and depth scaled by 4$\times$.  But when the best-known branch
predictors are afforded exponentially more resources, 80\% of this
opportunity remains untapped. New approaches are needed to extract this 
performance, which lies in just a handful of static branches in each application.

For the first time, branch prediction poses an attractive deployment scenario for machine learning
(ML). Gains in branch predictors over past
decades have balanced strict BPU constraints with the need for high
accuracy on thousands of static branches at a time.  Solutions have
favored simple, lightweight pattern-matching
%that can be replicated many times
~\cite{cbp2016, seznec_cbp16, lintarsa2019}, while comparatively
powerful, yet expensive ML models such as deep neural networks,
support vector machines, and random forests have been left
unexplored.  The large IPC opportunity that remains, and its
concentration in a few H2Ps that resist existing techiques
leads us to pursue ML models that implement more sophisticated pattern
matching within the BPU.

We propose ML-driven \emph{helper predictors} that operate alongside a
baseline predictor to boost accuracy for individual H2Ps.  This report
provides a tutorial developing convolutional neural network (CNN)
helpers to improve pattern matching on the same global history data
used by existing branch predictors.  We show how convolutional filters
better tolerate distortions in history data caused by control
structures with variable iteration counts.  We then train CNN
helpers with 2-bit weights and translate their inference procedure
into a small number of table lookups that meet BPU constraints.
Finally, we evaluate CNN helpers on applications traced over multiple
inputs to establish that gains hold in future executions.  At full
precision, CNN helpers reduce mispredictions by an average of 36.6\% on 47\% of H2Ps in SPECint 2017; our implementable design improves 24\% of H2Ps by 14.4\%.    

We adopt a deployment scenario in which helpers are trained to runtime
data \emph{offline} and uploaded to the BPU in future application
executions to generate predictions
\emph{online}~\cite{ravi2017charstar, tarsaisca2019}. This approach
amortizes training over the lifetime of a device and across devices
that run the same application, e.g. in a datacenter.  The result is an
application-specific IPC boost that requires no access to source
code or painstaking expert analysis.  Given a rich set of ML helper
predictors, we envision an optimization service that automatically
fits the best helper to each H2P and packages those that maximize IPC
as application metadata.  CNN helpers solve one source of systematic
misprediction, and we intend this report as a blueprint for the
development of other ML-driven helpers.

\section{Mispredictions Due to Variable Iteration Control Structures}
\label{sec:mispred}
We motivate CNN helpers by showing one class of H2P that arises due to
control structures with data-dependent iteration counts.  Two
examples, one illustrative and the other drawn from \texttt{deepsjeng}
in SPEC 2017 demonstrate that this common motif causes \emph{positional
variations} in data available to the BPU.  Such distortions confound
state-of-the-art predictors that rely on exact sequence matching or
positional correlations, but are tolerated by
convolutional filters. These examples are predicated on the following%% \footnote{See our companion paper~\cite{lintarsa2019} for additional details.}
:

\begin{itemize}[leftmargin=*]
\item We consider conditional branches only;
  % (\textbf{Taken} or \textbf{Not Taken});
\item When a branch is fetched, its \textbf{global history} is the
  sequence of instruction pointer values (IPs) and the 
  directions of branches executed leading up to the
  current instruction; %%  Global history is a proxy for an application's
  %% execution path and the data most commonly modeled in BPUs;
\item \textbf{TAGE-SC-L} is the state-of-the-art branch
  predictor~\cite{seznec_cbp16}. It conditions each prediction on the
  longest recognized subsequence of global history by approximating
  Partial Pattern Matching (PPM)~\cite{cleary84}. It is implemented by
  hashing global history subsequences into tagged tables. Table
  entries hold a saturating counter that tallies previously observed
  directions, and can be thresholded to make a prediction. TAGE-SC-L
  also implements a loop predictor, arbitrating between this and the
  longest-matching PPM predictions using a \emph{statistical
    corrector}, itself a perceptron;
\item The \textbf{perceptron \emph{predictor}} (distinct from the
  statistical corrector above) is an alternative to PPM predictors that
  trains weights for each global history position, isolating
  directions correlated to the current
  prediction~\cite{jimenez16}.  This mechanism filters noisy data that
  affects TAGE-SC-L's exact-match hash lookups, but requires
  positional weights to be stored and retrieved for many branches;
\item We identify as \textbf{H2Ps} any branch predicted to $<$99\%
  accuracy under TAGE-SC-L, and which generates at least 1,000 mispredictions
  per 30M-instruction SimPoint~\cite{lintarsa2019, john2018}.
\end{itemize}
\medskip
\label{sec:toy_example}
\begin{figure}[t!]
\begin{lstlisting}[frame=single, language=C, commentstyle=\color{blue},
  basicstyle=\ttfamily\tiny,numbers=left, numbersep=4pt, escapechar=@,
  xleftmargin=2pt, label={lst:toy}, captionpos=b, caption={A simple C
    function illustrates how common program structures cause systematic branch
    mispredictions.}]
int f(int k, int *uvec, int *vvec) {
  int val1 = 0;
  int val2 = 0;

  if (uvec[k] % 3 > 0) /*Data-Dependent Branch*/     @\label{lstline:h2p-1}@
    val1 += 1;

  for(int j = 0; j < (vvec[k]); j++) @\label{lstline:loop}@
    if (vvec[j] % 2 > 0) val2 += vvec[j];        @\label{lstline:uncorrelated_branch}@
           
  if (val1 > 0)         /*H2P-1*/      @\label{lstline:h2p-2}@
    return val2; 
  return 0;
}
\end{lstlisting}
\vspace{-0.75cm}
\end{figure}
\noindent \textbf{Illustrative Example --} Listing~\ref{lst:toy}
showcases an H2P (Line~\ref{lstline:h2p-2}, H2P-1) whose global history is affected by a loop with a variable
iteration count.  %% It contains a data-dependent branch at
%% Line~\ref{lstline:h2p-1} and an H2P at Line~\ref{lstline:h2p-2}
%% (H2P-1). 
H2P-1 is exactly correlated to the data-dependent
branch at Line~\ref{lstline:h2p-1}, and both branches are biased to be
taken 33\% of the time when \texttt{uvec}'s values are uniformly
distributed.  Crucially, they are separated by a loop whose iteration
count depends on data.  This code contains a simple, stable
relationship that predicts H2P-1---with no additional information on
data values, these two branches should be predicted to 66\% and 100\%
accuracy, respectively.

When a simple program calls function $f$ repeatedly with random
inputs, H2P-1's global histories exhibit significant variations. The
loop at Line~\ref{lstline:loop} injects different numbers of
uncorrelated branches into history data, causing the position of the
predictive data-dependent branch to change relative to H2P-1.  This
positional variation explodes the number of unique histories memorized
by a PPM predictor and breaks perceptron predictors that require
positional consistency.  Consequently, TAGE-SC-L predicts H2P-1
with 68\% accuracy, storing statistics in table entries
corresponding to all tracked subsequence lengths, while reusing few
for prediction.  Training a perceptron on H2P-1's global history
gives a similar 69\% accuracy.  In contrast, a CNN helper predicts H2P-1 with 100\%
accuracy (see Section~\ref{sec:Tutorial}).

\label{sec:deepsjeng_example}
%% BlackPawns Arr1
%% WhitePawns Arr2
%% BlackRooks Arr3
%% WhiteRooks Arr4
%% BlackKnights Arr5
%% WhiteKnights Arr6
%% BlackBishops Arr7
%% Arr8 Arr8
%% BlackQueens Arr9
%% WhiteQueens Arr10
%% BlackKing Arr11
%% WhiteKing Arr12

%% KnightMoves Arr13
%% KingMoves   Arr14
%% PawnAttacksBlack Arr15
%% PawnAttacksWhite Arr16
%% PawnMovesBlack Arr17
%% PawnMovesWhite Arr18

%% Mask Arr19
%% InvMask Arr20
                
%% DiagMaska1h8 Arr21
%% DiagMaska8h1 Arr22
%% FileMask Arr23
%% RankMask Arr24
%% AboveMask Arr25
%% BelowMask Arr26
%% LeftMask Arr27
%% RightMask Arr28
%% RookMask Arr29
%% BishopMask Arr30
%% QueenMask Arr31
%% CastleMask Arr32
%% FileUpMask Arr33
%% FileDownMask Arr34
%% WhiteKingSide Var1
%% WhiteQueenSide Var2
%% BlackKingSide  Var3
%% BlackQueenSide Var4
%% KingSafetyMask  Arr35
%% KingSafetyMask1 Arr36

%% WhiteStrongSquareMask Var5
%% BlackStrongSquareMask Var6

%% WhiteSqMask Var7
%% BlackSqMask Var8

%% KSMask Var9
%% QSMask Var10
                                 
%% KingFilesMask Arr37
%% KingPressureMask Arr38
%% KingPressureMask1 Arr39
%% CenterMask Var11
%% SpaceMask Arr40
%% material Arr41

%% MAT_PAWN                CONST_1
%% MAT_KNIGHT              CONST_2
%% MAT_BISHOP              CONST_3
%% MAT_ROOK                CONST_4
%% MAT_QUEEN               CONST_5
%% WHITE CONST_6
%% BLACK CONST_7
\begin{table}[t!]
\begin{lstlisting}[language=C, basicstyle=\ttfamily\tiny,numbers=left,
		   numbersep=4pt, xleftmargin=2pt, xrightmargin=0pt, 
		   frame=single, linewidth=\columnwidth, escapechar=@,
		   linebackgroundcolor={
					 \ifnum\value{lstnumber}=39
					 \color{red!60}
					 \fi
					 %% \ifnum\value{lstnumber}=40
					 %% \color{red!60}
					 %% \fi
                                         \ifnum\value{lstnumber}=3
					 \color{orange!35}
					 \fi
                                         \ifnum\value{lstnumber}=4
					 \color{orange!35}
					 \fi
                                         \ifnum\value{lstnumber}=5
					 \color{orange!35}
					 \fi
                                         \ifnum\value{lstnumber}=11
					 \color{blue!20}
					 \fi
                                         %% \ifnum\value{lstnumber}=25
					 %% \color{yellow!60}
					 %% \fi
                                         \ifnum\value{lstnumber}=25
					 \color{yellow!60}
					 \fi
                                         \ifnum\value{lstnumber}=14
					 \color{orange!35}
					 \fi
                                         \ifnum\value{lstnumber}=13
					 \color{orange!35}
					 \fi
                                         \ifnum\value{lstnumber}=16
					 \color{orange!35}
					 \fi
                                         \ifnum\value{lstnumber}=19
					 \color{orange!35}
					 \fi
                                         \ifnum\value{lstnumber}=22
					 \color{orange!35}
					 \fi
                                         \ifnum\value{lstnumber}=28
					 \color{orange!35}
					 \fi
                                         \ifnum\value{lstnumber}=31
					 \color{orange!35}
					 \fi
                                         \ifnum\value{lstnumber}=36
					 \color{orange!35}
					 \fi
                                         %% \ifnum\value{lstnumber}=42
					 %% \color{orange!35}
					 %% \fi
                                         %% \ifnum\value{lstnumber}=45
					 %% \color{orange!35}
					 %% \fi
                                         %% \ifnum\value{lstnumber}=48
					 %% \color{orange!35}
					 %% \fi
                                         %% \ifnum\value{lstnumber}=51
					 %% \color{orange!35}
					 %% \fi
                                         %% \ifnum\value{lstnumber}=54
					 %% \color{orange!35}
					 %% \fi
                                         \ifnum\value{lstnumber}=40
					 \color{orange!35}
					 \fi
		   }]		     
typeA_t func1(typeB_t *a, const int b) {
  // ...    
  if (Arr23[v4] & v2) { v1 |= (v2 & func1b(A2, b)); } @\label{lstline:dsj_b1}@
  if (Arr21[b] & v3) { v1 |= (v3 & func1c(A2, b)); }  @\label{lstline:dsj_b2}@
  if (Arr22[b] & v3) { v1 |= (v3 & func1d(A2, b)); }  @\label{lstline:dsj_b3}@
  return v1;
}

int func2(typeB_t *a, const int b, const int c, const int c, const int d) {
  // ...
  lc = func1(a, c);                                   @\label{lstline:dsj_lc}@
  // ... 
  while (lc) {                                        @\label{lstline:dsj_b4}@
    if (v8 == CONST_6) {                              @\label{lstline:dsj_b5}@
      v1 = Arr2 & lc;
      if (v1) { v7 = CONST_1; }                       @\label{lstline:dsj_b6}@
      else {
        v1 = Arr6 & lc;
        if (v1) { v7 = CONST_2; }                     @\label{lstline:dsj_b7}@
        else {
          v1 = Arr8 & lc;
          if (v1) { v7 = CONST_3; }                   @\label{lstline:dsj_b8}@
          else {
            v1 = Arr4 & lc;
            if (v1) { v7 = CONST_4; }   /* CORRELATED BRANCH A */              @\label{lstline:dsj_b9}@
            else {
              v1 = Arr10 & lc;
              if (v1) { v7 = CONST_5; }               @\label{lstline:dsj_b10}@
              else {
                v1 = Arr12 & lc;
                if (!v1) { break; }                   @\label{lstline:dsj_b11}@
                v7 = CONST_0;
              }
    // ...    
    } else {
        //...6 more nested correlated branches (not shown)...
    }
    // ...        
    if (v4 & Arr29[c]) { lc |= func2a(a, c) & v4; }    /** H2P BRANCH **/ @\label{lstline:dsj_h2p}@        
    if (v5 & Arr30[c]) { lc |= func2b(a, c) & v5; }  @\label{lstline:dsj_b19}@
    lc &= a->A2;    	                             @\label{lstline:dsj_lc2}@
    // ...	
  } 
  // ...
}
\end{lstlisting}
\captionof{lstlisting}{An H2P in
  \texttt{deepsjeng}
  (Line~\ref{lstline:dsj_h2p}, red) has many correlated branches
  (orange, yellow).  The loop condition is set on
  Line~\ref{lstline:dsj_lc} (blue) but modified over time
  (Lines~\ref{lstline:dsj_h2p}--\ref{lstline:dsj_lc2}), causing
  correlated branches to shift position in the H2P's global history.}
\label{lst:deepsjeng}
\vspace{-0.25cm}
\end{table}

\begin{figure}[t!]
\centering
\includegraphics[clip,width=\columnwidth]{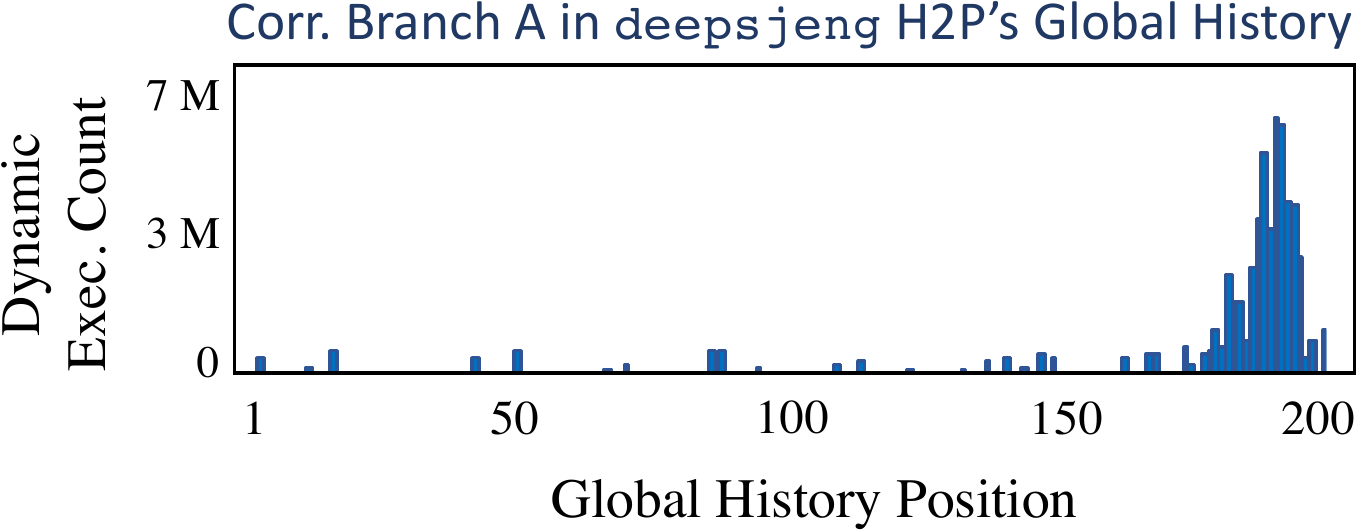}
\caption{Histogram of Correlated Branch A's position 
  (Listing~\ref{lst:deepsjeng}, Line~\ref{lstline:dsj_b9}, yellow), which
  exhibits positional variations in an H2P's global histories. 
   History position 200 is the most recent.}
\label{fig:ip_pos_histogram}
\vspace{-0.5cm}
\end{figure}
\smallskip
\noindent \textbf{SPECint 2017 deepsjeng --} Positional variations
appear in the wild in SPECint 2017, for example in \texttt{deepsjeng}.  Listing~\ref{lst:deepsjeng}
shows a code fragment
%\footnote{This codehas been redacted such that control
%  flow is apparent to conform to the SPEC license.} 
 from the \texttt{deepsjeng} source containing an H2P
branch at Line~\ref{lstline:dsj_h2p} (red).
Lines~\ref{lstline:dsj_b1}-\ref{lstline:dsj_b3}, \ref{lstline:dsj_b4},
\ref{lstline:dsj_b5}, \ref{lstline:dsj_b6}, \ref{lstline:dsj_b7},
\ref{lstline:dsj_b8}, \ref{lstline:dsj_b9}, \ref{lstline:dsj_b10},
\ref{lstline:dsj_b11}, and \ref{lstline:dsj_b19} (orange and yellow)
show correlated branches in the H2P's global history. They all reside within a loop
conditioned on variable~\texttt{lc}, whose value is initialized
at Line~\ref{lstline:dsj_lc} but modified over iterations
 (Lines~\ref{lstline:dsj_h2p}-\ref{lstline:dsj_lc2}).  
As a result, the loop iteration count is variable and 
correlated branches shift position in the H2P's global history. 

This is evident on examination of just one of the correlated branches,
Correlated Branch A (Line~\ref{lstline:dsj_b9}, yellow).
Fig.~\ref{fig:ip_pos_histogram} shows that the distribution of its
position in the H2P's histories roughly spans the most recent 25
positions.  Increasing global history subsequence length, an
optimization made when scaling TAGE-SC-L from 8KB to 64 KB, does not
directly address this variation, which is the root cause of this H2P.
Positional variations are also exhibited by all other (orange)
correlated branches in Listing~\ref{lst:deepsjeng}.  As a result, we find 
that TAGE-SC-L predicts the H2P on Line~\ref{lstline:dsj_h2p} with
just 93.8\% accuracy, while a CNN will predict it with 100\%
accuracy.

\section{A CNN Global History Model}
\label{sec:Tutorial}
%% We next provide a step-by-step walkthrough of how a small
%% convolutional neural network can model branch history data. CNNs are
%% distinguished from perceptrons in their use of (1) multiple layers, to
%% represent hierarchies of different features extracted from the data,
%% and (2) convolution, to provide shift-tolerant pattern matching at
%% lower layers.

%% We begin with the two-layer network shown in
%% Fig.~\ref{fig:ExampleCNN}, and first illustrate how a trained CNN
%% performs inference, i.e., pattern matching and prediction generation.
%% We then describe the training procedure.  Ultimately, when applied to
%% Listing~\ref{lst:toy}, this network predicts H2P-1 with 100\% accuracy
%% using fixed-length global histories of 200 prior branches. For
%% simplicity, throughout this paper we assume that one CNN is trained
%% per H2P.
%\begin{figure}[t]
%\centering
%\includegraphics[clip,width=0.90\columnwidth]{./figures/TwoLayerBP-CNN.pdf}
%\caption{A two-layer CNN gives 100\% accuracy for H2P-1 from Listing~\ref{lst:toy}.}
%\label{fig:ExampleCNN}
%\end{figure}
%% To show how a CNN predicts H2P-1, we walk through the forward pass of
%% the two-layer network in Listing~\ref{fig:ExampleCNN}, i.e. prediction-generation
%% assuming a trained network, then discuss training.
To show how a CNN predicts H2P-1, we first walk through the forward pass of the
two-layer CNN in Listing~\ref{fig:ExampleCNN} to produce a prediction, 
and then describe how it is trained. We initially use the
full network representation, but in Section~\ref{sec:online_inf}
translate it into a mechanism that meets BPU constraints.
\vspace{-0.25cm}
\subsection{Encoding History Data}
\label{sec:encode}

\begin{figure}[t]
  \centering
  \subfloat{
          \includegraphics[width=0.95\linewidth]{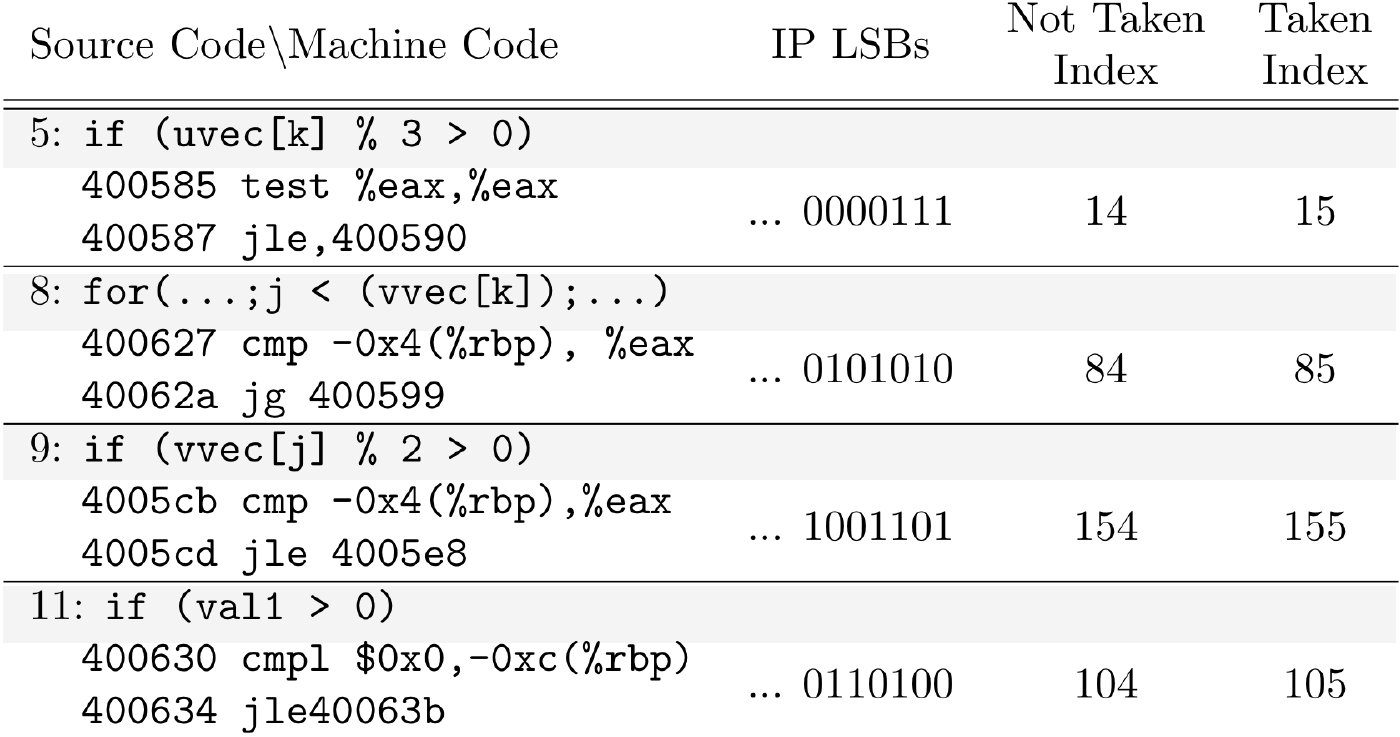}
          \label{fig:EmbedIdx}
  }
  \vspace{-0.4cm}
  \\
  \subfloat{
  \includegraphics[width=0.95\linewidth]{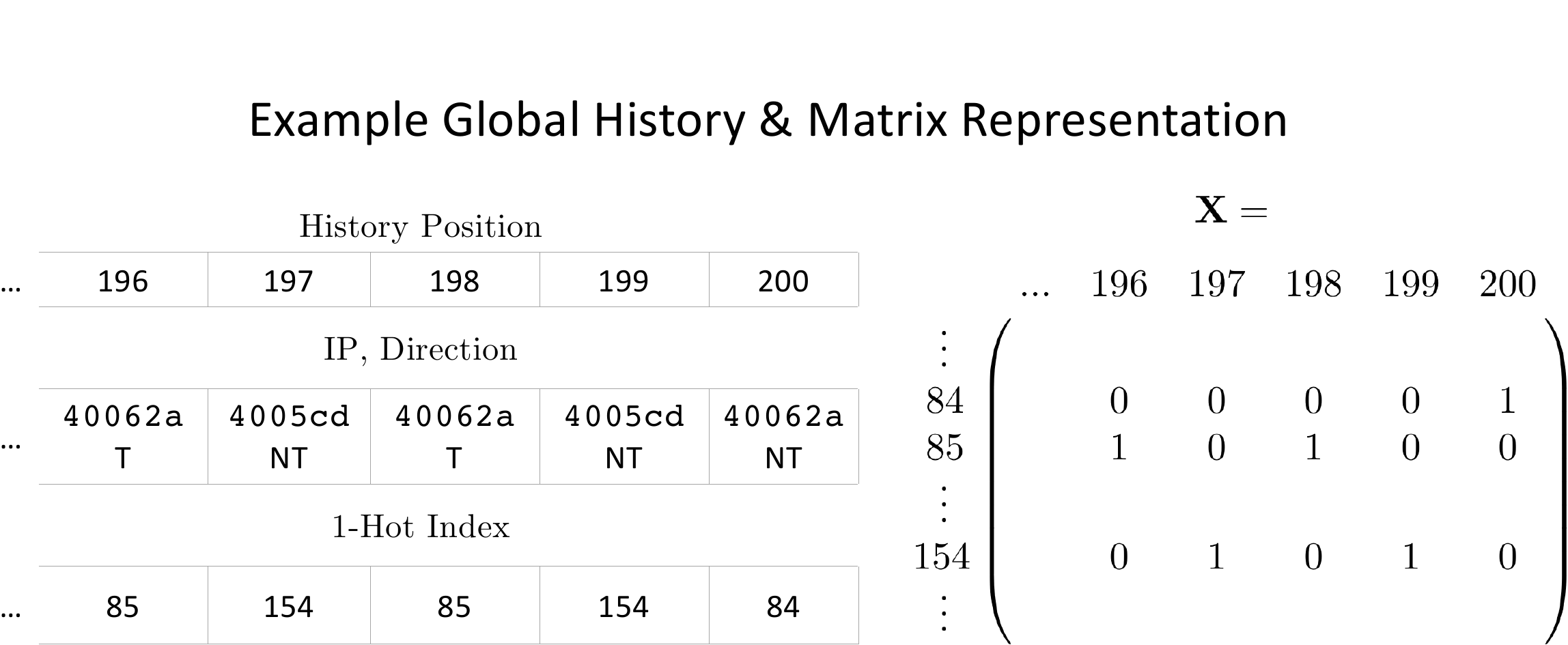}
  \label{fig:EmbedMat}
}
%% \vspace{0.25cm}
\caption{A CNN is fed H2P-1's global history as a matrix
  of 1-hot columns with 1's in
  indices $((IP\ll1) + Dir) \& (2^{p}-1)$.}
\label{fig:Embed}
%\vspace{-0.25cm}
\end{figure}

Given a dynamic instance of an H2P, we convert its global history
sequence of $\langle$IP, direction$\rangle$ tuples into an algebraic
vector representation. IPs are discrete and take on a large number of
possible values, so we use a hash function to index into a ``1-hot"
vector, which contains a one at the index position and zeros
otherwise.  For example, setting vector dimension to $2^p = 256$, we
map each tuple to $[0, ..., 255]$ by concatenating the
observed-direction bit onto the $p-1$ LSBs of the IP: $((IP \ll 1) +
Dir) \& (2^{p}-1)$.  This process is shown in Fig.~\ref{fig:Embed} for
H2P-1. Four branches from Listing~\ref{lst:toy} are shown alongside
their IP values, observed directions, and the indices used to generate
1-hot vectors.  We concatenate column vectors to form a global history
matrix $X$, which is input to the CNN.

Though 1-hot history matrices appear costly in terms of storage, we choose
this encoding because matrices can be replaced on-BPU
with direct-mapped table lookups (Section~\ref{sec:online_inf}). 
Our experiments show that our CNNs perform well
with as few as seven LSBs from each IP, making them
agnostic to an application's base virtual address.
To ensure history encodings behave consistently 
across executions, we set $p=8$.

\begin{table}[t!]
\begin{lstlisting}[frame=single, language=Python, commentstyle=\color{blue},
  basicstyle=\ttfamily\tiny,numbers=left, numbersep=4pt, escapechar=@,
  xleftmargin=2pt, label={fig:ExampleCNN}, captionpos=b, caption={A simple CNN gives
  100\% accuracy for H2P-1.}]
class TwoLayerFullPrecisionBPCNN(chainer.Chain):

  # Network definition
  def __init__(self, tableSize=256, numFilters=2, historyLen=200):
    super(TwoLayerFullPrecisionBPCNN, self).__init__()
    with self.init_scope():
      # Embed via identity matrix:
      self.E  = cupy.identity(tableSize, dtype=cupy.float32)
    
      # Convolution layer, 1D in effect; Linear layer:
      self.c1 = L.Convolution2D(16,  numFilters, (1, 1))
      self.l2 = L.Linear(historyLen*numFilters, 1)

  # Forward pass
  def __call__(self, x):
    # Embed by expanding sequence of ((IP << 7) + dir) & (255) 
    # integers into 1-hot history matrix during training:    
    xFull = cupy.rollaxis( cupy.take(self.E, x.data, axis=0), 3, 1)
    
    h1  = self.c1(xFull)
    h1a = tanh(h1)
    h2  = self.l2(h1a)

    return F.sigmoid(self.h2)
    
    
\end{lstlisting}
\vspace{-0.25cm}
\end{table}

\subsection{Layer 1: Convolutional Correlation}
\label{sec:layer1_conv_corr_detect}

\begin{figure}
  %\centering
  \subfloat{
    \includegraphics[width=0.91\linewidth]{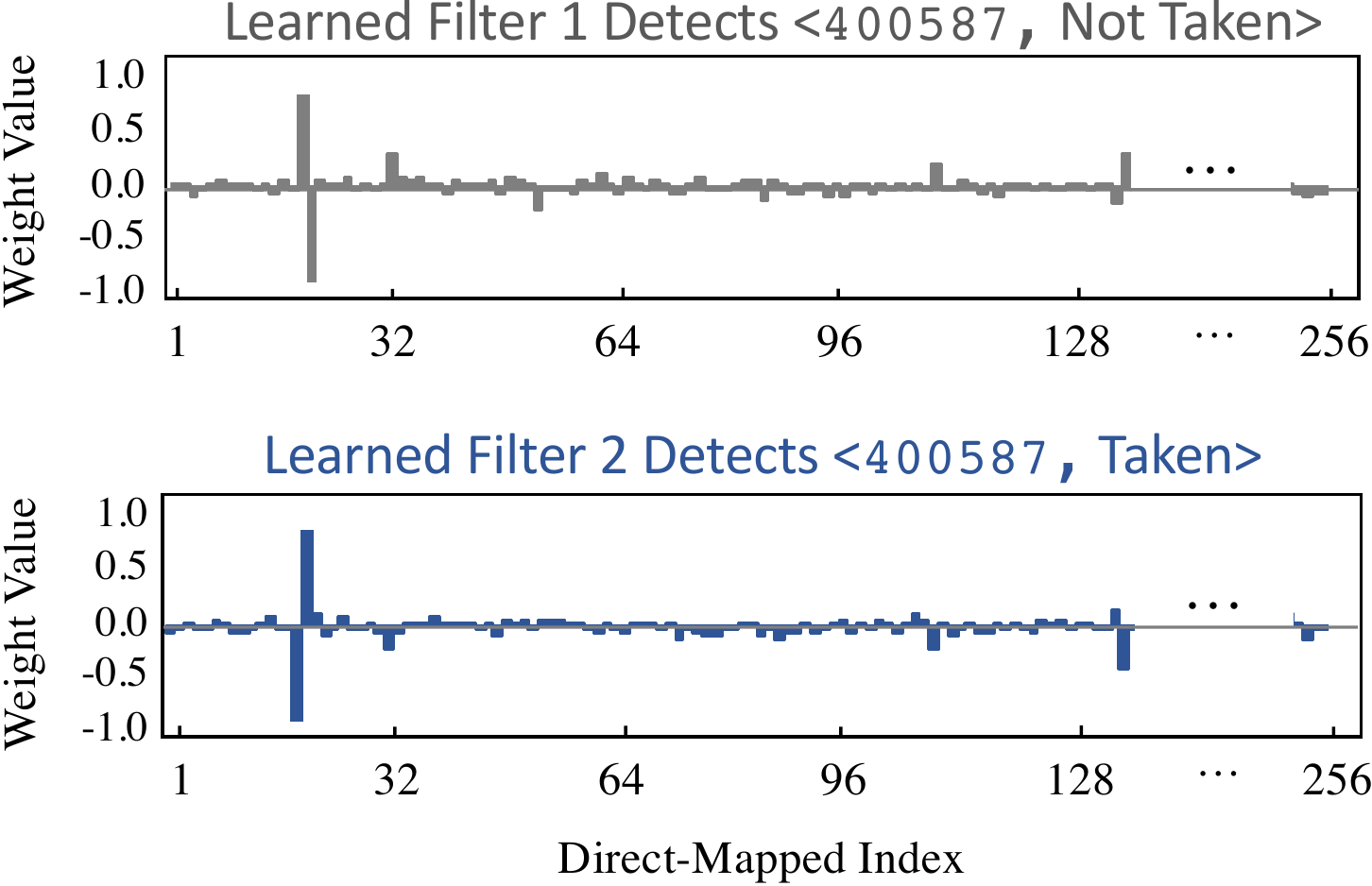}
%    \label{fig:FilterDefs}
  }\\
  \centering
  \subfloat{
    \includegraphics[width=0.80\linewidth]{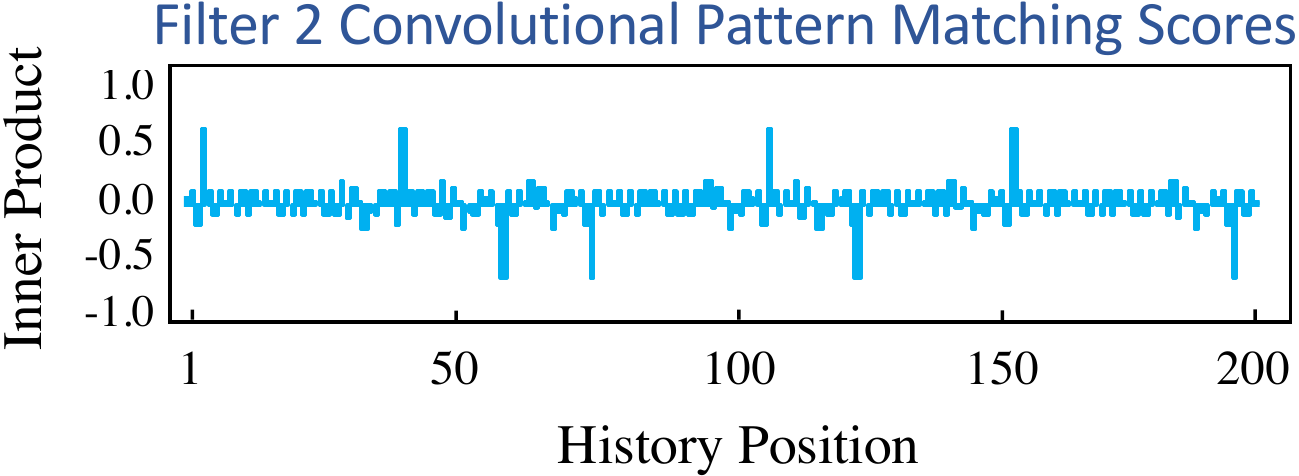}
%    \label{fig:FilterScores}
  }
  \caption{(Top 2) 1-wide convolutional filters trained on
    H2P-1's history learn large weights at indices of the data-dependent branch;
    (Bot.) convolutional pattern
    matching gives large values at positions where that branch appears in H2P-1's global
    history.}
  \label{fig:ConvFilters}
  \vspace{-0.25cm}
\end{figure}

CNNs perform pattern matching using inner-product computations between
a data vector $x$ and a weight vector $w$, also called a \emph{filter}, 
with an optional bias term $b$. A similar computation is used in perceptron predictors, however our CNNs differ
by performing the same filter matches at every
history position, and by also matching on IPs.
\begin{equation} \label{eq1}
y = \sum_{i}w_ix_i + b.
\end{equation}
To illustrate,  we instantiate our CNN with two filters and plot their
values in 
Fig.~\ref{fig:ConvFilters} (Top) after training on history 
matrices and observed directions for H2P-1.
We see that Filter 1 learns a large positive
weight at index 14, aligning to correlated
branch \texttt{0x400587} being not-taken, while Filter 2 exhibits a
large weight at index 15 for the same IP being taken.  
Small weights adjust for branches that are
biased in H2P-1's history, though magnitudes are
negligible in comparison.  

Evaluating Eq.~\ref{eq1} for each filter against 
each column of the history matrix produces $200*2$ 
inner product scores for history length 200.
Fig.~\ref{fig:ConvFilters} (Bottom) shows the 200 scores computed from Filter 2. We
call \texttt{f()} from a loop, so H2P-1's global history also contains
stale appearances of the correlated data-dependent branch, and each
produces a large filter response.

\subsection{Layer 2: Positional Prediction}
\begin{figure}
  \includegraphics[clip,width=0.90\columnwidth]{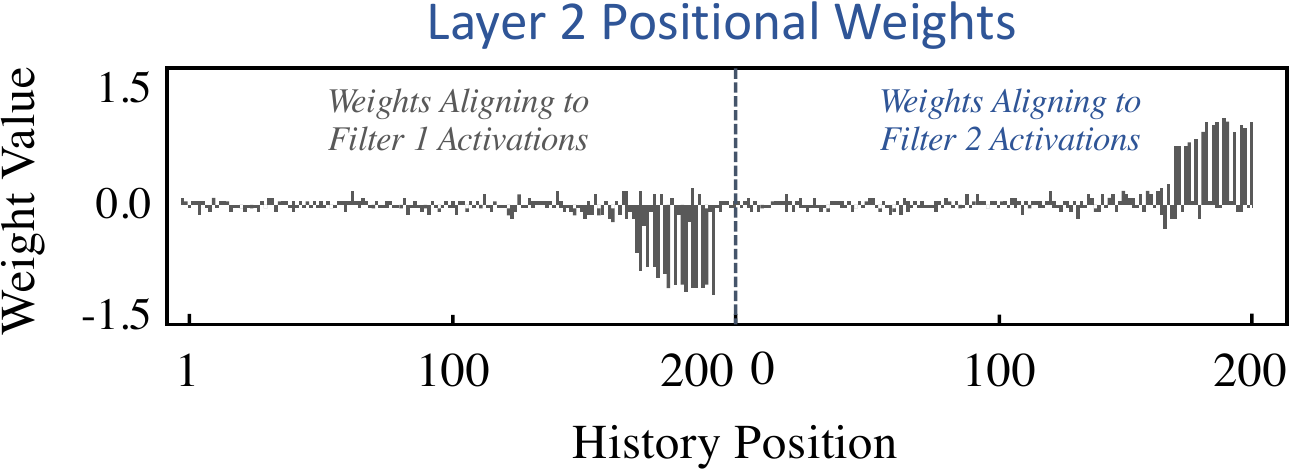}
  \caption{Layer 2 filter weights represent how much each history
    position contributes to the final prediction.}
  \label{fig:PredictionWeights}
  \vspace{-0.25cm}
\end{figure}
Scores computed in the convolution layer above are
passed to a perceptron-like \emph{linear layer}, which contains a
single filter that is matched against the output of Layer 1.
The trained weights of Layer 2's filter are shown in
Fig.~\ref{fig:PredictionWeights}.  Near-zero weights
damp positions beyond the most recent
30, filtering out stale appearances of IP \texttt{0x400587}.  %Since Filter 1 and 2 represent
%``not-taken'' and ``taken'', respectively, their positional responses
%in Layer 2 are anti-correlated, as expected.
%
%%where we see that a positive detection for Filter 1 pushes the prediction toward Not-Taken by yielding a negative score, while a detection from Filter 2 pushes the prediction toward Taken. Furthermore, 
Eq.~\ref{eq1} is applied once at Layer 2 using its filter
and the Layer 1 scores as inputs. 
%The result
%predicts ``taken'' if greater than zero and ``not-taken'' otherwise.  %% For training purposes, the CNN applies a
%% sigmoid function to this last layer, however at prediction time a
%% simple threshold is sufficient (see Section~\ref{sec:online_inf}).
%% Outputs from the sigmoid predict H2P-1 with 100\% accuracy.

\subsection{Stacking the Layers Together}
The result of Layer 2's pattern matching operation 
predicts ``taken'' if greater than zero and ``not-taken'' otherwise.
The two layers of this CNN handle different aspects of predicting
H2P-1's direction:  the first layer is position-agnostic and is
designed to identify which $\langle$IP, direction$\rangle$ tuples in a
branch history correlate highly with the H2P's direction; the second
layer is designed to identify which positions in a
branch history contribute most to the prediction.  
%As a branch
%history is fed into the CNN, Layer 1 discards those branches that
%are unimportant, and Layer 2 discards branches that
%appear in positions of the branch history that play no role in the
%prediction.  
The combined filtering action of these stacked layers
allows the CNN to precisely latch onto the predictive signal in H2P-1's 
histories, even as it shifts position---\emph{it is this mechanism, 
the result of stacking convolutional and linear layers, that gives
our CNNs a pattern matching advantage over PPM and perceptron predictors}. 

%% The two layers of this network handle different aspects 
%% By following a position-agnostic layer with position-specific layer,
%% the CNN is capable of filtering away unimportant branches at Layer
%% 1.  The second layer weights enable CNN predictors to focus on specific
%% portions of a branch's history.  Here, this capability enables the CNN
%% helper to discard data beyond position 30.
%% CNNs are succesful in many
%% domains because stacked modeling layers can be trained to capture
%% intricate structures from data.  For branch prediction, CNNs model a
%% variety of predictive program structures from runtime statistics,
%% requiring no access to source code.

\subsection{Offline Training}
\label{sec:training}
%
%As discussed in Section~\ref{sec:background:deployment}, we intend CNN
%helper predictors for offline-training/online-inference because we
%target deployments where repeated executions drive one-time training costs to zero.  As
%a result, our helpers do not have strict training time
%requirements, and can use the most advanced techniques and
%hardware available.  Here, we provide a brief overview of training.
The training dataset for a CNN helper consists of history matrices of an H2P alongside its 
observed directions, which we collect using the Pin binary instrumentation tool~\cite{luk2005pin}. 
We train networks using Chainer~\cite{tokui2015chainer}
%\&
%PyTorch~\cite{paszke2017automatic}
%open-source packages,
and find, for the CNN configurations used in
Section~\ref{sec:eval}, 5,000 history-matrix/H2P-direction pairs sampled
uniformly from runtime data are sufficient to converge
in 40 epochs using the Adam optimizer~\cite{kingma2014adam}.

\section{On-BPU Inference with 2-Bit CNNs}
\label{sec:online_inf}
To deploy our CNN helper in a BPU, we train networks with 2-bit weights
and show that they need only modest on-chip storage and bitwise-parallel logic at prediction time. 
CNNs provide strong pattern recognition even when their weights are constrained to values in $\{+1,
0, -1\}$~\cite{courbariaux16,rastegari16}, allowing logical operations
to replace arithmetic during inference.  %We take
%advantage of this property so that our CNN helpers can fit in a BPU
%that is both storage- and latency-constrained.

%To deploy our CNN helper in a BPU that is constrained in 
%storage and latency, we train networks with 2-bit weights
%and show that they need only modest on-chip storage and bitwise-parallel logical operations at prediction time. 
%% The literature has shown that CNNs provide strong pattern recognition even when weights are constrained 
%% to take on values in $\{+1, 0, -1\}$~\cite{courbariaux16,rastegari16}, allowing us to replace arithmetic
%% operations with logical ones during inference.
%inner products require bitwise logic, popcount,
%and subtract.

\begin{figure}[t!]
  \includegraphics[clip,width=0.92\columnwidth]{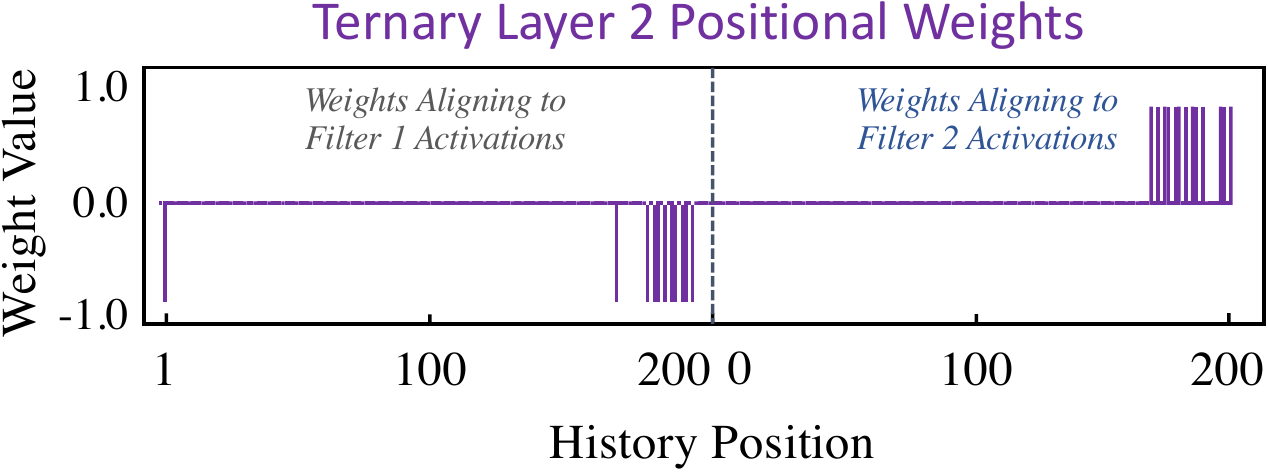}
  \caption{2-bit CNN helpers lose fidelity encoding the magnitude of each position's
    contribution to predictions, but accurately detect $\langle$IP,
    direction $\rangle$ tuples despite positional variations.}
  \label{fig:TernaryPredictionWeights}
  \vspace{-0.25cm}  
\end{figure}

Following Courbariaux \emph{et al.}~\cite{courbariaux16}, we impose
low-precision constraints during training by
clipping weights to $[-1, +1]$, normalizing activations, and quantizing during
forward CNN computations (Listing~\ref{fig:TernaryNetwork}). 
%When 2-bit weights are stored as sign and value bits, 
We train the resulting \emph{ternary} CNN helper for H2P-1
on the same training data.  Fig.~\ref{fig:TernaryPredictionWeights} shows ternary Layer 2 weights. Compared to the full-precision
weights in Fig.~\ref{fig:PredictionWeights}, 
quantized weights lose accuracy encoding the magnitude of each
position's contribution to predictions, but still detect correlated $\langle$IP,
direction$\rangle$ tuples and damp stale data. This ternary CNN helper 
yields 98\% accuracy for H2P-1.
%\begin{figure}[t]
%  \includegraphics[clip,width=0.90\columnwidth]{figures/TwoLayerTernaryBP-CNN.pdf}
%  \caption{CNNs can be constrained to have low-precision weights by adding
%  quantization and normalization layers.}
%  \label{fig:TernaryNetwork}
%\end{figure}

\begin{table}[t!]
\begin{lstlisting}[frame=single, language=Python, commentstyle=\color{blue},
  basicstyle=\ttfamily\tiny,numbers=left, numbersep=4pt, escapechar=@,
  xleftmargin=2pt, label={fig:TernaryNetwork}, captionpos=b, caption={CNNs can be constrained
  to have 2-bit weights during training by adding quantization and normalization
  steps.}]
class TwoLayerTernaryBPCNN(chainer.Chain):

  # Network definition
  def __init__(self, tableSize=256, numFilters=2, historyLen=200):
    super(TwoLayerTernaryBPCNN, self).__init__()    
    with self.init_scope():
      # Embed via identity matrix:
      self.E  = cupy.identity(tableSize, dtype=cupy.float32)
    
      # Ternary convolution layer, now with batch normalization
      self.c1 = TernConv2D(tableSize, numFilters, kernel=(1,1))
      self.b1 = BatchNormalization(numFilters)
            
      # Ternary linear Layer, now with batch normalization
      self.l2 = TernLinear(numFilters*historyLen, 1, nobias=True)
      self.b2 = L.BatchNormalization(1)

  # Forward pass
  def __call__(self, x):
    # Embed by expanding sequence of ((IP << 7) + dir) & (255) 
    # integers into 1-hot history matrix during training:    
    xFull = cupy.rollaxis( cupy.take(self.E, x.data, axis=0), 3, 1)
    
    h1  = self.c1(xFull)
    h1n = self.b1(h1)
    h1t = F.TernQuant(h1n)

    h2  = self.l2(h1t)
    h2n = self.b2(h2)
    
    return F.sigmoid(self.h2n)
\end{lstlisting}
\vspace{-0.25cm}
\end{table}

%\begin{figure}[t]
%  \centering
%  \includegraphics[clip,width=1\columnwidth]{./figures/PredGenDiagram.pdf}
%  \caption{2-bit CNN helper inference computations are pipelined as
%    data arrives.}
%  \label{fig:pred_gens}
%\end{figure}

%\subsubsection{On-BPU Ternary Inference Implementation}
%\label{sec:onbpu}
Not only does accuracy remain high, but ternary CNN inference for our
network can also be made far more efficient than its full-precision
counterpart based on the following observations:
%
%These simple mechanisms exploit the following observations:
%\begin{itemize}%[leftmargin=*]

\medskip
\noindent \textbf{Multiplication by a 1-hot vector yields a
  scalar.} The inner product of a 1-hot input vector and 
  a filter is the filter value that aligns
  to the input's sole non-zero.  We therefore sidestep
  history matrices completely on-BPU by indexing 
  $\langle$IP, direction$\rangle$ tuples into a table of filter values. 
  Subsequent normalization and quantization steps can also be folded into this table, since they produce 
  a 2-bit value from each possible filter value after training. We precompute this half 
  of the inference computation for any input by populating a lookup table as follows:
  For $m$
  filters of length $2^p$, denoted $W = [w_1, \ldots, w_m]$; indices
  $i \in [1, 2^{p}], j \in [1, m]$;
  learned parameters $\mu_{1j}, \sigma_{1j}, \gamma_{1j}, \beta_{1j}$ from a
  normalization layer that transforms data according to $\hat{y_j} =
  (y_j-\mu_{1j})(\gamma_{1j}/\sigma_{1j}) + \beta_{1j}$; and quantization bins defined
  over the ranges $[-1, -q]$, $[-q, +q]$, $[q, 1]$, we populate a $2^p \times
  m \times 2$-bit table $\mathcal{T}$ as:
  \begin{equation} \label{layer1eq}
    \begin{split}
      & \mathcal{T}[i,j] = \begin{cases}
        01, \text{ if } w_{ij} < \frac{-\beta_{1j}}{\gamma_{1j}}\sigma_{1j} + \mu_{1j} - q\\
        11, \text{ if } w_{ij} > \frac{-\beta_{1j}}{\gamma_{1j}}\sigma_{1j} + \mu_{1j} + q\\
        00, \text{ otherwise.}
      \end{cases}%, \\
%      & \quad\quad\quad\quad\text{for } i \in [1, 2^{p}], j \in [1, m]
    \end{split}
  \end{equation} 
  \noindent $q$ defines quantization buckets
  for ternary CNN weights~\cite{courbariaux16,rastegari16}; we set
  $q=0.8$ but note its value may be learned~\cite{zhu2016trained}.
  
  \medskip
\noindent \textbf{1-wide convolutions can be computed as soon as
  history tuples are available.}  When applying convolutions of width
1, filter responses for each history position are independent of their
neighbors.  This allows us to retrieve the Layer 1 responses well
before an H2P is encountered, as $\langle$IP, direction$\rangle$
tuples become available.  Whenever a prior branch's direction is
predicted, the corresponding Layer 1 responses are retrieved from
$\mathcal{T}$ and pushed into a FIFO buffer. When an H2P is fetched
and a CNN prediction needed, Layer 1 outputs are already available in
the FIFO buffer.
  
  \medskip  
\noindent \textbf{Ternary inner products require only bitwise parallel
  logic, popcount, and subtraction.} At prediction time, we evaluate
Layer 2 and its normalization layer.  This entails an inner product
between the FIFO buffer's contents and ternary weights, scaling and
shifting the resulting integer value by learned normalization
parameters, and comparing to 0 to give a 
``taken'' or ``not-taken'' prediction. We implement the ternary inner product
as:
  \begin{equation}
    \small
    \begin{split}
      P = \text{\texttt{popcount}}(\neg(L1_S \land L2_S) \& (L1_V \& L2_V)) - \\
      \text{\texttt{popcount}}((L1_S \land L2_S) \& (L1_V \& L2_V))\\
    \end{split}
  \end{equation}
  \noindent where $L1_{S}$ and $L1_{V}$ are the sign and value bits of
  the FIFO buffer, respectively, and $L2_{S}$ and $L2_{V}$ contain
  those for the Layer 2 filter. We apply the inverse of normalization to 0 to solve
  for a threshold $t$, above which we predict taken:
  \begin{equation}\label{eqn:pred}
    \begin{split}
      \text{Pred} = \begin{cases}
        1, \text{ if } P > t, \textrm{where } t = \frac{-\sigma_2}{\gamma_2}\beta_2 - \mu_2 \\
        0, \text{ otherwise}
      \end{cases}
    \end{split}
  \end{equation}
%\end{itemize}

On-BPU CNN inference thus consists of two steps, defined by
Algorithms~\ref{ternUpdateAlg} and~\ref{ternInfAlg}.  The first is a
table lookup to update the FIFO buffer of Layer 1 filter responses
whenever any dynamic conditional branch is fetched.  The second is a
ternary inner product between the FIFO buffer and Layer 2 filter
when the H2P is fetched and a prediction needed.  Any time a
branch is mispredicted, the CPU is rolled back to that instruction, and
wrong-path entries are simply shifted off the FIFO buffer.

%\noindent We note that this inference procedure is of similar complexity to those 
%used by TAGE at prediction time, when the history is hashed, multiple values retrieved from tables, and results compared. 

\begin{algorithm}[t!]
  \caption{FIFO Update}\label{ternUpdateAlg}
  \begin{algorithmic}[1]
  \small
    \Require{ $\langle$IP, Dir$\rangle$ of most recent branch, 
     $\mathcal{T}$ as in Eq.~\ref{layer1eq}; L1, a $200 \times m \times 2$ bit buffer; 
     $p=8$}.
%    \Statex
    \Function{CNNHistoryUpdate}{IP, Dir}
        \State i $\gets$ (IP $\ll$ 1 + Dir) \& $(2^{p}-1)$
        \State L1 $\gets$ (L1 $\ll$ 2*m) $\mid$ $\mathcal{T}$[i, 0...2*m]
    \EndFunction
  \end{algorithmic}
\end{algorithm}

%% \vspace{-0.5cm}
\begin{algorithm}[t!]
  \caption{2-Bit CNN Branch Prediction}\label{ternInfAlg}
  \small
  \begin{algorithmic}[1]
    \Require{L1, a $200 \times m \times 2$ bit buffer filled per Alg.~\ref{ternUpdateAlg};
     L2, a $200 \times m \times 2$ bit buffer of Layer 2 weights; 
     $t$ as in Eq.~\ref{eqn:pred}}
 %   \Statex
    \Function{CNNPredict}{}
    \State  s\_bits $\gets$ L1$_S$ $\land$ L2$_S$
    \State  v\_bits $\gets$ L1$_V$ $\&$ L2$_V$
    \State  P $\gets$ \texttt{popcount(}$\neg$(s\_bits \& v\_bits)\texttt{)}
    \Statex \makebox[1.2cm]{}- \texttt{popcount(}s\_bits \& v\_bits\texttt{)}
%      \State $f[i] \gets L_{1}[h_{IP}[i] \ll 1 + h_{Dir}[i]]$
    
    \State \textbf{return} (P $>$ t)
    \EndFunction
  \end{algorithmic}
\end{algorithm}

\subsection{On-BPU Storage and Latency}

To install a CNN helper in a BPU, we must store four components:
(1) a $2^p \times m \times 2$-bit table to hold Layer 1 filter
responses; (2) a $\texttt{historyLen} \times m \times 2$-bit FIFO
buffer to hold convolution results; (3) a $\texttt{historyLen} \times m
\times 2$-bit buffer to hold the Layer 2 weights; (4) a buffer to
hold the precomputed integer threshold.  
Our network, with $p=8$, $m=2$, and $\texttt{historyLen}=200$, 
requires 336 bytes per helper.  For 
$m=32$, storage is 5.2KB.

While a full layout of our CNN helper is beyond the scope of this report, we 
can compare the relative latency between ternary CNN inference and TAGE-SC-L 
by analyzing the computation graphs of their prediction procedures (Table~\ref{ref:inf_complexity}). 
For example, in Algorithm ~\ref{ternInfAlg}, we are able to compute
Lines 2--3 in parallel, compute 
(s\_bits \& v\_bits) on Line 4
  and $\neg$ operations serially, the \texttt{popcount}s in
  parallel, and finally the subtract and comparison serially. Predictions from a
  2-bit CNN helper thus require six serial computations. The
  bottleneck computation is \texttt{popcount}, which requires a
  13 or 15 stage circuit depending on $m$~\cite{ramanarayanan20082}.
In contrast, TAGE-SC-L 8KB and 64KB require 34 and 32 serial
  computations, respectively (TAGE-SC-L 8KB uses more
  complex hashing). Their bottleneck computations are 
  back-to-back lookups to 4k and 8k entry tables, depending on predictor.  This comparison
  shows that a ternary CNN helper requires a similar number
  of computation steps to existing predictors.

\section{CNN Helper Gains \& Reusability}
\label{sec:eval}

\begin{table*}
\resizebox{\textwidth}{!}{%
\renewcommand{\arraystretch}{0.9}
\begin{tabular}{llcc|cc|cc|cc}
\hline
\hline
\multicolumn{2}{c|}{\multirow{2}{*}{\begin{tabular}[c]{@{}c@{}}SPECint2017\\ Benchmark\end{tabular}}} & \multicolumn{1}{c}{\multirow{2}{*}{\begin{tabular}[c]{@{}c@{}}\# Training \\ Folds\end{tabular}}} & \multicolumn{1}{c|}{\multirow{2}{*}{\begin{tabular}[c]{@{}c@{}}\# H2Ps \\ (All Phases)\end{tabular}}} & \multicolumn{2}{l|}{FP-CNN with TAGE 8KB Baseline}                                                                                                                              & \multicolumn{2}{l|}{TP-CNN with TAGE 8KB Baseline}                                                                                                                             & \multicolumn{2}{l}{FP-CNN, Gains Beyond TAGE 64KB}   \\ \cline{5-10} 
\multicolumn{2}{c|}{}                                                                                 & \multicolumn{1}{c}{}                                                                               &\multicolumn{1}{c|}{} & \multicolumn{1}{l|}{\% Winners} & \multicolumn{1}{c|}{\begin{tabular}[c]{@{}c@{}}Mispred. \\ Red. per H2P\end{tabular}} & \multicolumn{1}{c|}{\% Winners} & \multicolumn{1}{c|}{\begin{tabular}[c]{@{}c@{}}Mispred.\\ Red. per H2P\end{tabular}} & \multicolumn{1}{l|}{\% Winners} & \multicolumn{1}{c}{\begin{tabular}[c]{@{}c@{}}Mispred.\\ Red. per H2P\end{tabular}} \\ \hline \hline
\multicolumn{2}{l|}{600.perlbench\_s}                                                                 & \multicolumn{1}{c}{4}                                                                             & 16 & 51\%                            & 63.2\%                                                                                                                         & 18\%                            & 26.6\%                                                                                                                      & 4\%                             & 8.2\%                                                                                                                         \\
\multicolumn{2}{l|}{605.mcf\_s}                                                                       & \multicolumn{1}{c}{8}                                                                             & 20 & 55\%                            & 44.8\%                                                                                                                                & 28\%                            & 27.9\%                                                                                                                     & 35\%                            & 19.3\%                                                                                                                  \\
\multicolumn{2}{l|}{620.omnetpp\_s}                                                                   & \multicolumn{1}{c}{5}                                                                             & 28 & 71\%                            & 33.6\%                                                                                                                                & 30\%                            & 16.3\%                                                                                                                   & 24\%                            & 11.2\%                                                                                                                      \\
\multicolumn{2}{l|}{623.xalancbmk\_s}                                                                 & \multicolumn{1}{c}{4}                                                                             &  8 & 39\%                            & 27.4\%                                                                                                                                & 0\%                             & 0.0\%                                                                                                                      & 23\%                            & 12.8\%                                                                                                                     \\
\multicolumn{2}{l|}{625.x264\_s}                                                                      & \multicolumn{1}{c}{14}                                                                            &  7 & 44\%                            & 16.8\%                                                                                                                                & 35\%                            & 12.0\%                                                                                                                     & 33\%                            & 12.2\%                                                                                                                    \\
\multicolumn{2}{l|}{631.deepsjeng\_s}                                                                 & \multicolumn{1}{c}{12}                                                                            & 49 & 56\%                            & 31.2\%                                                                                                                                & 24\%                            & 10.0\%                                                                                                                     & 12\%                            & 15.3\%                                                                                                                    \\
\multicolumn{2}{l|}{641.leela\_s}                                                                     & \multicolumn{1}{c}{10}                                                                            & 68 & 68\%                            & 40.7\%                                                                                                                               & 44\%                            & 15.3\%                                                                                                                      & 41\%                            & 19.7\%                                                                                                                      \\
\multicolumn{2}{l|}{645.exchange2\_s}                                                                 & \multicolumn{1}{c}{5}                                                                             & 19 & 9\%                             & 46.5\%                                                                                                                               & 4\%                             & 6.0\%                                                                                                                       & 0\%                             & 0.0\%                                                                                                                      \\
\multicolumn{2}{l|}{657.xz\_s}                                                                        & \multicolumn{1}{c}{5}                                                                             & 50 & 28\%                            & 25.2\%                                                                                                                                & 29\%                            & 15.4\%                                                                                                                     & 15\%                            & 12.3\%                                                                                                                      \\ \hline
\multicolumn{2}{l|}{MEAN}                                                                              & \multicolumn{1}{c}{7.3}                                                                          & 29 & 47\%                            & 36.6\%                                                                                                                               & 24\%                            & 14.4\%                                                                                                                    & 21\%                            & 12.3\%                                                                                                                                            \\ \hline \hline
\end{tabular}}
\caption{CNN Helpers reusably improve accuracy for a large portion of H2Ps. Gains for 21\% of 
H2Ps are beyond the capabilities of TAGE-SC-L when scaled by 8x.}
\label{tab:cnn_results}
\vspace{-0.25cm}
\end{table*}

We demonstrate CNN helpers on SPECint 2017 and assess reusability with the 
dataset of~\cite{lintarsa2019}, which traces each benchmark over multiple inputs%% \footnote{We note that the
%% Alberta Workloads for SPEC2017 provide an alternative multi-input dataset~\cite{amaral2018alberta}}
. 
For each benchmark, we screen for H2Ps using TAGE-SC-L 8KB 
as the baseline predictor in the Championship Branch Prediction
2016 simulator~\cite{cbp2016}, and train a
CNN helper for any H2P appearing in 3 or more application inputs (i.e. workloads)
to support $k$-fold cross-validation. We train on data
from the entirety of a single workload and report performance averaged across all 
held-out workloads; this constitutes one fold, and we average 
all possible folds to compute the \emph{expected}
gains in future executions, assuming we train on data
from an arbitrary execution.  \emph{Training on one workload and testing on
the hold-outs demonstrates the reusability of our CNNs.}
We evaluate full-precision CNN Helpers (FP-CNN) as a limit study
and ternary CNNs (TP-CNN). For both, we use a history length of 200, 
encode 7 bits of each IP and 1 direction bit, and
32 Layer 1 filters.

\begin{table}[]
\renewcommand{\arraystretch}{0.8}
\setlength{\tabcolsep}{2pt}
\resizebox{0.47\textwidth}{!}{%
\begin{tabular}{rcccc}
\hline \hline
\multicolumn{1}{c}{}                                                                             & \multicolumn{4}{c}{Prediction Generation Complexity}                                                                                                                                                           \\ \hline
\multicolumn{1}{c|}{}                                                                            & {\begingroup \renewcommand{\arraystretch}{1.15} \begin{tabular}[c]{@{}c@{}}TAGE\\ 8 KB\end{tabular}   \endgroup }                             & {\begingroup \renewcommand{\arraystretch}{1.15} \begin{tabular}[c]{@{}c@{}}TAGE\\ 64 KB\end{tabular}     \endgroup   }                        & {\begingroup \renewcommand{\arraystretch}{1.15} \begin{tabular}[c]{@{}c@{}}TP-CNN\\ 8 filter\end{tabular} \endgroup } & {\begingroup \renewcommand{\arraystretch}{1.15} \begin{tabular}[c]{@{}c@{}}TP-CNN\\ 32 filter\end{tabular} \endgroup } \\ \hline \hline
\multicolumn{1}{r|}{\begingroup \renewcommand{\arraystretch}{0.8} \begin{tabular}[r]{@{}r@{}}\# Serial Computations\end{tabular} \endgroup }                                              & \multicolumn{1}{c}{\textbf{34}}           & \multicolumn{1}{c}{\textbf{32}}           & \textbf{6}                                                & \textbf{6}                                                 \\ 
\multicolumn{1}{r|}{\begin{tabular}[c]{@{}c@{}}\# Serial Tbl. Lkups. \end{tabular}} & \multicolumn{1}{c}{\textbf{2}} & \multicolumn{1}{c}{\textbf{2}} & \textbf{0}                                                & \textbf{0}                                                 \\
%\multicolumn{1}{r|}{ \begingroup \renewcommand{\arraystretch}{1.6} Total Bit Ops \endgroup}                                                    & { \begingroup \renewcommand{\arraystretch}{1.2} \multicolumn{1}{c}{4,839}  \endgroup }          & { \begingroup \renewcommand{\arraystretch}{1.2} \multicolumn{1}{c}{3,716}   \endgroup }              & {\begingroup \renewcommand{\arraystretch}{1.2} 9,610   \endgroup}                                                   & {\begingroup \renewcommand{\arraystretch}{1.2}  38,410   \endgroup }                                                  \\ 
%\multicolumn{1}{r|}{\begingroup \renewcommand{\arraystretch}{1.2} \begin{tabular}[r]{@{}r@{}}Total Bit Ops. \end{tabular} \endgroup}                                               & \multicolumn{1}{c}{ \begingroup \renewcommand{\arraystretch}{1.2} \begin{tabular}[c]{@{}c@{}}4,839 \end{tabular} \endgroup }   & \multicolumn{1}{c}{ \begingroup \renewcommand{\arraystretch}{1.2} \begin{tabular}[c]{@{}c@{}}3,716 \end{tabular} \endgroup }   & { \begingroup \renewcommand{\arraystretch}{1.2} \begin{tabular}[c]{@{}c@{}}9,600\end{tabular}   \endgroup }                                             & { \begingroup \renewcommand{\arraystretch}{1.2} \begin{tabular}[c]{@{}c@{}} 38,400 \end{tabular}  \endgroup }                                       \\
%\multicolumn{1}{r|}{}& \multicolumn{1}{c}{} & \multicolumn{1}{c}{} & & \\ 
\multicolumn{1}{r|}{\begingroup \renewcommand{\arraystretch}{0.8} \begin{tabular}[r]{@{}r@{}}Latency Limiting\\ Computation\end{tabular} \endgroup}                                               & \multicolumn{1}{c}{ \begingroup \renewcommand{\arraystretch}{1.0} \begin{tabular}[c]{@{}c@{}}2$\times$ lookup,\\ 4k-entry \\table\end{tabular} \endgroup }   & \multicolumn{1}{c}{ \begingroup \renewcommand{\arraystretch}{1.0} \begin{tabular}[c]{@{}c@{}}2$\times$ lookup,\\ 8k-entry \\table\end{tabular} \endgroup }   & { \begingroup \renewcommand{\arraystretch}{1.0} \begin{tabular}[c]{@{}c@{}}\texttt{popcount} \\ \emph{(13} \emph{stage} \\ \emph{circuit)}\end{tabular}   \endgroup }                                             & { \begingroup \renewcommand{\arraystretch}{1.0} \begin{tabular}[c]{@{}c@{}}\texttt{popcount}\\ \emph{(15} \emph{stage}\\ \emph{circuit)}\end{tabular}  \endgroup }                                       \\ \hline \hline
\end{tabular}}
\caption{At prediction time, CNN Helpers perform a ternary inner product followed by popcount.}
\label{ref:inf_complexity}
\vspace{-0.25cm}
\end{table}

Table~\ref{tab:cnn_results} breaks out the portion of CNN helper
predictors that improved H2P accuracy (\% Winners) by benchmark,
alongside the accuracy improvement per H2P (\% Reduction
in Mispredictions).  On average, the FP-CNN shows that pattern matching
with tolerance for positional variations improves accuracy on 47\% of H2Ps
by an average 36.6\% reduction in mispredictions, \emph{reusably across workloads}. 
When we use FP-CNN helpers \emph{and} scale TAGE-SC-L to 64KB, we still find additional gains---21\%
of H2Ps improve by 12.3\% on average, improving gains in mispredictions-per-kilo-instruction (MPKI) 
from 21.2\% to 22.3\% over the baseline. This shows one example when improved pattern
matching provides a fundamental advantage over scaling existing algorithms. 

TP-CNN helpers improve 24\% of H2Ps by 14.4\% on average, capturing
roughly half the gain of FP-CNNs.  Given that quantizing Layer 2 weights
in TP-CNN (Fig.~\ref{fig:TernaryPredictionWeights}) tempers the 
positional precedence captured by a full-precision Layer 2
(Fig.~\ref{fig:PredictionWeights}), this comparison shows
that arbitrating with potentially stale data is also an important contributor to prediction accuracy.

%% Given the tradeoffs illustrated in Figs.~\ref{fig:PredictionWeights}
%%  and~\ref{fig:TernaryPredictionWeights}, this shows that the positional precedence captured
%%  by a full-precision Layer 2 but lost to quantization in TP-CNN affects many predictions, e.g., due to stale data. %% This
 %% particular implementable TP-CNN design captures approximately half of the improvement of FP-CNN.

%Fig.~\ref{fig:ipc} reports IPC improvements across SPECint 2017
%for various BPU configurations, measured against a SkyLake core configuration with
%TAGE-SC-L 8KB as the baseline predictor. At all configurations, 
%both TP-CNN and FP-CNN provide additional, complementary IPC gains beyond TAGE-SC-L 64KB. Table~\ref{ref:ipc_table} 
%reports MPKI alongside IPC showing that, while helpers produce seemingly small
%improvements in MPKI, their impact on IPC is outsized, consistent with our data
%in @@@ showing that H2Ps have an outsized impact on IPC. At 1x pipeline scale, 
%the difference in performance impact of CNN helpers is over 
%4:1 for the same reduction in MPKI.

%% Our objectives are (1) to show that CNN helper predictors are reusable
%% across application executions with different inputs, and (2) to offer
%% an algorithmic alternative to existing techniques, while providing
%% novel, attractive design tradeoffs.  From this, we hope to spur further
%% exploitation of off-BPU training, and deeper exploration of advanced
%% machine learning techniques to target H2Ps.

\section{Directions for Future ML Helpers}
This paper details how a two-layer CNN reduces systematic branch
mispredictions caused by positional-variations in global history data.
We demonstrate a path to deployment that (1) meets on-BPU constraints
for prediction generation, and (2) can amortize iterative batch
training through reuse across application executions.  Several
natural future directions exist:
\begin{enumerate}[leftmargin=*]
\item CNNs provide an expressive pattern matching framework and
  support rapid experimentation; exploring topologies,
  e.g. to learn predictive multi-IP subsequences or extract patterns
  from arbitrarily long global histories using recurrence can address
  different causes of misprediction;
\item The gap between FP-CNN and TP-CNN shows the need for alternative
  on-BPU designs, that, e.g., integrate dependent branch IPs
  identified by a CNN into lightweight predictors.  In such a design,
  ML models act as an automated analysis tool, rather than an on-BPU
  predictor directly;
\item Feeding additional data such as register values into ML models
  may boost prediction accuracy for data-dependent branches. In this manner, 
  an ML model acts as a approximate value predictor, possibly
  exploiting idle multiply-accumulate cycles in the core. 
\end{enumerate}
These avenues and others will provide fruitful ground for machine learning
in branch predictor development. 
%
%
%
%We presented a new approach to branch prediction that uses convolutional
%neural networks to extract additional IPC from
%the same data available to traditional predictors.  We showed that CNN helpers are
%practical for generating predictions online in the BPU when
%we employ low-precision, ternary weight CNNs. Most importantly, CNN helper predictors
%sustain their gains across application executions.  We can therefore
%relax the prevailing assumption that branch statistics \emph{must} be
%learned online in the BPU, and consider offline training
%mechanisms to open up new regions of
%the BPU design space to advanced machine learning methods. Further exploration
%of both other CNN topologies and other ML algorithms are likely to bear fruit for 
%branch predictors, and we believe machine learning 
%will be a centerpiece of branch prediction
%going forward.  
%% This allows for a major relaxation of a restrictive
%% assumption common in the literature: branch statistics can be learned
%% off-BPU. This paper lays the groundwork for new research directions
%% that harvest the gains of sophisticated machine learning advances to
%% improve CPU performance.

%% Our objectives are (1) to show that CNN helper predictors are reusable
%% across application executions with different inputs, and (2) to offer
%% an algorithmic alternative to existing techniques, while providing
%% novel, attractive design tradeoffs.  From this, we hope to spur further
%% exploitation of off-BPU training, and deeper exploration of advanced
%% machine learning techniques to target H2Ps.

\bibliographystyle{unsrt}
\bibliography{ref}

\begin{thebibliography}{10}

\bibitem{fog2018microarchitecture}
A~Fog.
\newblock The microarchitecture of intel, amd, and via cpus.
\newblock {\em An Optimization Guide for Assembly Programmers and Compiler
  Makers. Copenhagen University College of Engineering}, 2018.

\bibitem{seznec_cbp16}
A~Seznec.
\newblock {TAGE-SC-L Branch Predictors Again}.
\newblock In {\em Proc. 5th Championship on Branch Prediction}, 2016.

\bibitem{lintarsa2019}
C-K Lin and SJ~Tarsa.
\newblock {Branch Prediction is Not a Solved Problem: Measurements,
  Opportunities, and Future Directions}.
\newblock {\em arXiv:1906.08170}, 2019.

\bibitem{cbp2016}
{CBP-5 Kit}.
\newblock In {\em Proc. 5th Championship on Branch Prediction}, 2016.

\bibitem{ravi2017charstar}
GS~Ravi and MH~Lipasti.
\newblock {CHARSTAR: Clock Hierarchy Aware Resource Scaling in Tiled
  Architectures}.
\newblock {\em ACM SIGARCH}, 2017.

\bibitem{tarsaisca2019}
SJ~Tarsa, RBR Chowdhury, J~Sebot, GN~Chinya, J~Gaur, K~Sankaranarayanan, C-K
  Lin, R~Chappell, R~Singhal, and H~Wang.
\newblock {Practical Post-Silicon CPU Adaptation Using Machine Learning}.
\newblock In {\em ISCA}, 2019.

\bibitem{cleary84}
J~Cleary and I~Witten.
\newblock {Data Compression Using Adaptive Coding and Partial String Matching}.
\newblock {\em IEEE Trans Comms}, 1984.

\bibitem{jimenez16}
DA~Jim{\'e}nez.
\newblock {Multiperspective Perceptron Predictor}.
\newblock In {\em Proc. 5th Championship on Branch Prediction}, 2016.

\bibitem{john2018}
S~Song, Q~Wu, S~Flolid, and J~et~al Dean.
\newblock {Experiments with SPEC CPU 2017: Similarity, Balance, Phase Behavior
  and Simpoints}.
\newblock Technical report, TR-180515-01, Dept. of ECE, UT-Austin, 2018.

\bibitem{luk2005pin}
C-K Luk, R~Cohn, R~Muth, H~Patil, A~Klauser, G~Lowney, S~Wallace, VJ~Reddi, and
  K~Hazelwood.
\newblock {Pin: Building Customized Program Analysis Tools with Dynamic
  Instrumentation}.
\newblock 2005.

\bibitem{tokui2015chainer}
S~Tokui, K~Oono, S~Hido, and J~Clayton.
\newblock {Chainer: A Next-Generation Open Source Framework for Deep Learning}.
\newblock In {\em LearnSys}, 2015.

\bibitem{kingma2014adam}
D~Kingma and J~Ba.
\newblock {Adam: A Method for Stochastic Optimization}.
\newblock {\em arXiv:1412.6980}, 2014.

\bibitem{courbariaux16}
M~Courbariaux, I~Hubara, D~Soudry, R~El-Yaniv, and Y~Bengio.
\newblock {Binarized Neural Networks: Training Deep Neural Networks with
  Weights and Activations Constrained to +1 or -1}.
\newblock {\em arXiv:1602.02830}, 2016.

\bibitem{rastegari16}
M~Rastegari, V~Ordonez, J~Redmon, and A~Farhadi.
\newblock {XNOR-Net: Imagenet Classification Using Binary Convolutional Neural
  Networks}.
\newblock In {\em ECCV}, 2016.

\bibitem{zhu2016trained}
C~Zhu, S~Han, H~Mao, and WJ~Dally.
\newblock {Trained Ternary Quantization}.
\newblock {\em arXiv:1612.01064}, 2016.

\bibitem{ramanarayanan20082}
R~Ramanarayanan, S~Mathew, V~Erraguntla, R~Krishnamurthy, and S~Gueron.
\newblock {A 2.1Ghz 6.5mW 64-bit Unified Popcount/Bitscan Datapath Unit for
  65nm High-Performance Microprocessor Execution Cores}.
\newblock In {\em VLSID}, 2008.

\end{thebibliography}

\end{document}